\documentclass[aps,prl,twocolumn,showpacs]{revtex4}

\usepackage{graphicx}% Include figure files
\usepackage{ulem}

\begin{document}
\title{ Quantum size effect in Pb(100) films: the role of symmetry and implication for film growth}
\author{Dengke Yu$^1$, Matthias Scheffler$^1$, and  Mats Persson$^{1,2}$ }
\affiliation{$^1$Theory Department, Fritz-Haber-Institut der Max-Planck-Gesellschaft,
Faradayweg 4-6, D-14195, Berlin-Dahlem, Germany }
\affiliation{$^2$Department of Applied Physics, Chalmers University of
Technology, SE-412 96, G{\"o}teborg, Sweden}

%\tighten
\begin{abstract}
{We show from density-functional calculations that Pb(100) thin
films exhibit quantum size effect with a bilayer periodicity in
film energies, film relaxations, and work functions, which
originate from different symmetry of the stacking geometry of odd
and even layer films. The bilayer periodicity of
the film energy is argued to survive on a semiconductor substrate,
which should allow  the growth of ``magically'' thick even-layer
Pb(100) films. Furthermore, it is found that the quantum well
states in a simple metal film can be classified into
$\sigma$-bonded and $\pi$-bonded states, which quantize independently. }
\end{abstract}

\maketitle

The so-called ``electronic growth'' model of metallic thin film on
a semiconductor substrate has attracted much scientific and
technological interest~\cite{Zhang}. This model provides
conditions for the growth of atomically-flat thin films or
nano-structures, which are of significant technological interest.
The underlying physical mechanism behind ``electronic growth'' is
the quantum size effect (QSE) on the film energy introduced by
quantum well (QW) states in the film. In addition to the film
energy, the energy gain associated by the charge transfer
occurring at the film semiconductor interface needs also to be
considered to determine the film stability. The total energy of
the system may then result in ``critical'' thicknesses in which
the film is stable when the number of atomic layers is above or
below a critical number and ``magical'' film thicknesses in which
the film has a pronounced stability for certain number of atomic
layers. The ``electronic growth" model has been very successful in
understanding observed ``critical'' thicknesses of alkali metal
and noble metal overlayers grown on a GaAs surface~\cite{Zhang},
and the observed ``magical''thicknesses of Pb(111) films or
islands on various substrates\cite{Yeh,Chiang1,Roberto,Hup}.

The nature of observed QW states and the associated QSE in metal
film have so far been well understood from simple models based on
bulk band structure and Fermi surface
nesting\cite{Stiles,ChiangR,Qiu,Milun}. In general, QW states are
formed by the quantization of the valence electrons confined in
the perpendicular direction of the thin film, as described by the
phase accumulation model.  The energies of the QW states are then
obtained from the energies of the valence bulk bands at the
quantized wave vectors. This simple model has been widely used in
analysis of photoemission measurements of QW
states~\cite{ChiangR,Qiu,Milun}. The QSE on the film energy and
other physical properties arises from the consecutive occupation
of QW states with increasing film thickness. An enhanced density
of states at the Fermi surface by QW states is then obtained from
extremal spanning (nesting) vectors $\bf q$ of the Fermi surface
that are perpendicular to the film, resulting in QSE period
$\lambda_{QSE}=2\pi/q$ for the film energy. For a free-electron
model of the bulk bands, corresponding to a spherical Fermi
surface, $q = 2k_F$ where $k_F$ is the Fermi wave vector and
$\lambda_{QSE} = \pi/k_F$.

This simple model for QSE has been supported by first-principles,
electronic structure calculations for various systems. Schulte~\cite{Schu}
found that the electron densities, potentials, and work functions of a thin
jellium film exhibited QSE with a period $\pi/k_F$ in accordance
with a free-electron model. Feibelman~\cite{Feib} and
Batra {\sl et. al.}~\cite{Batra} found a QSE in the surface energy
and surface
relaxation of both A1(111) and Mg(0001) thin films. Wei and Chou\cite{Wei} found
that the QSE period for surface
energies and work functions of  thin Pb(111) films was consistent with a spanning vector
in the Pb bulk band structure in [111] direction. The observed oscillations of film
thickness and interlayer spacing had the same QSE
period~\cite{Yeh,Hup,relax1,relax2,relax3}. So far the calculated
and observed QSE periods can be simply understood from a
nesting vector of the bulk band structure along the direction perpendicular
to the film, which projects to the  center of the film Brillouin zone (FBZ).

In this letter we show from a density-functional theory (DFT)
study of thin Pb(100) films that the symmetry of the stacking
geometry of a thin film can have a profound influence on the QSE
through QW states at the FBZ boundary. In particular, the film
energy, relaxations and work functions exhibit a bilayer
periodicity that cannot be understood in terms of nesting vectors
of the Fermi surface along the $\rm \Gamma X$  direction in the
bulk BZ. This symmetry effect is a general phenomena for fcc(100)
(and bcc(100)) films but  should only be important for the film
energy when the material also has a large band gap around Fermi
level at the FBZ boundary.

We have carried out DFT calculations of the total energy,
geometric and electronic structure of freestanding Pb(100) films
using a pseudo-potential, plane-wave basis method within the
local-density approximation. Technical details can be found in
Ref. \cite{yu}. Spin-orbit coupling was not considered.

\begin{figure}
\includegraphics[width=8.5cm]{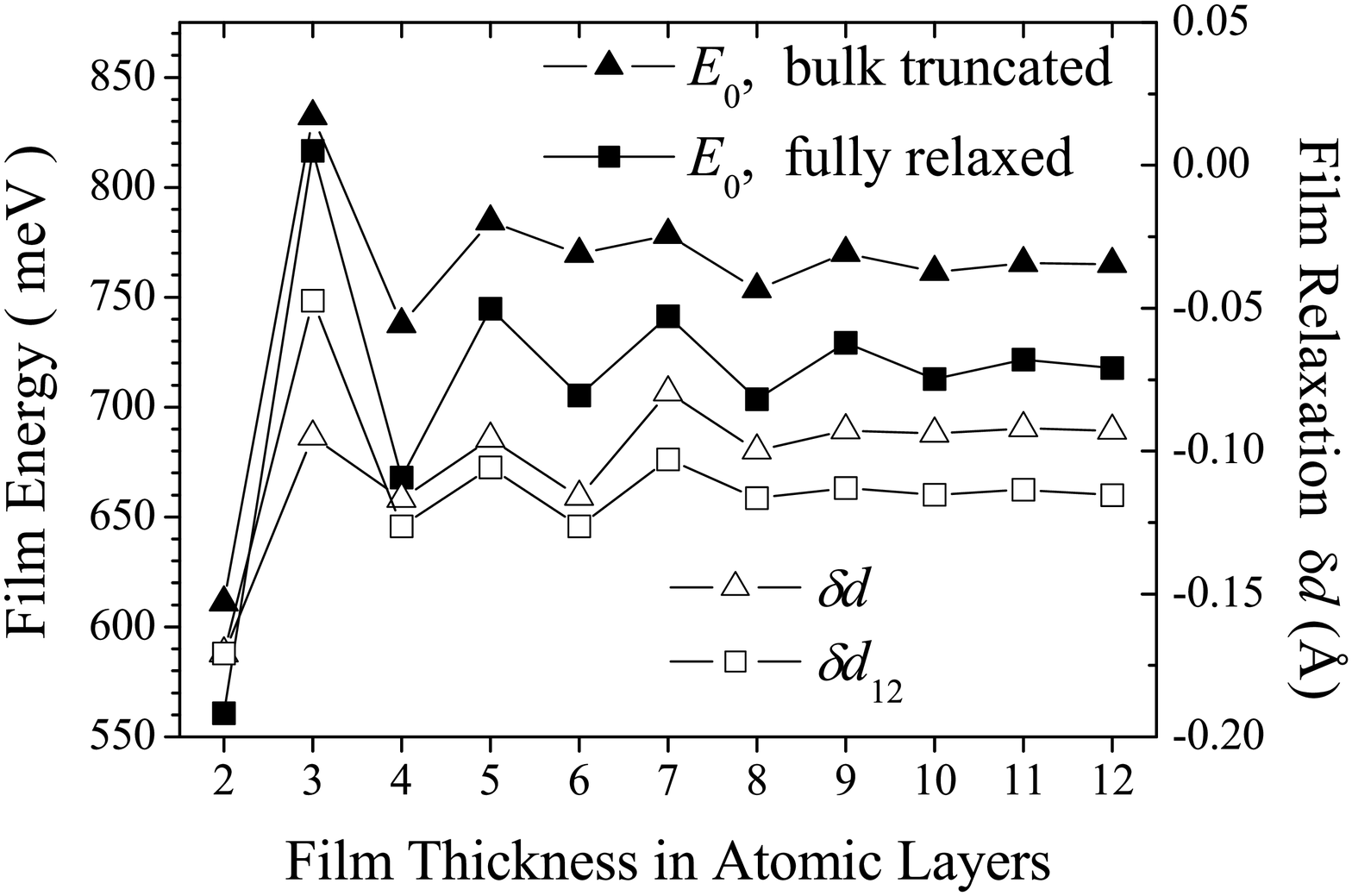}
\caption{\label{Fig:1}Film energy $E_0$ per unit cell (left axis) and
geometrical relaxations $\delta d$ and $\delta d_{12}$(right axis)
of free-standing Pb(100) films with respect to the number of
atomic layers.  $E_0$ of both the bulk truncated and
the fully relaxed film are shown.  $E_0$ is obtained by subtracting the bulk
contribution from the total energy of the film per  unit cell.
$\delta d$ and $\delta d_{12}$
are the changes of the total film thickness and the first
interlayer spacing from their truncated bulk values, respectively.
Note that the energy and the relaxations exhibit a characteristic
odd-even-layer alternation.  }
\end{figure}

As shown in Fig. 1, the calculated film energies and
relaxations of the free-standing, thin Pb(100) films exhibit
pronounced QSE. All these quantities exhibit a characteristic
damped odd-even-layer alternation. The film energies of the
relaxed films have a much more pronounced bilayer oscillation than
the bulk truncated film, showing that the QSE on the relaxations
contribute to the oscillations of the film energy in a cooperative
manner. Both the film thickness and the first interlayer spacing
show very similar oscillatory behavior. Also the work functions of
the relaxed films (not shown here) exhibit QSE and oscillate in an
anti-phase manner relative to the oscillations of the film energy
(i.e, work function maxima appear at the film energy minima).

A key issue is to understand why the QSE in thin Pb(100) films has
a bilayer period. We first show that this periodicity cannot
simply be understood in terms of  nesting vectors of the  Fermi
surface along the $\rm \Gamma X$ direction. The band structure around
the Fermi level is dominated by bands derived from atomic $p$
orbitals. Along the
$\rm \Gamma X$ direction, corresponding to the perpendicular direction
of the film, the Fermi surface has a nesting vector $q =
1.36\pi/a$ corresponding to $\lambda_{QSE}=2.9d_{100}$ where
$d_{100}$ is the inter-layer spacing. Thus a QSE period of about
three (100) layers is expected in contrast to the calculated
bilayer period for the QSE.

\begin{figure}
\includegraphics[width=8cm]{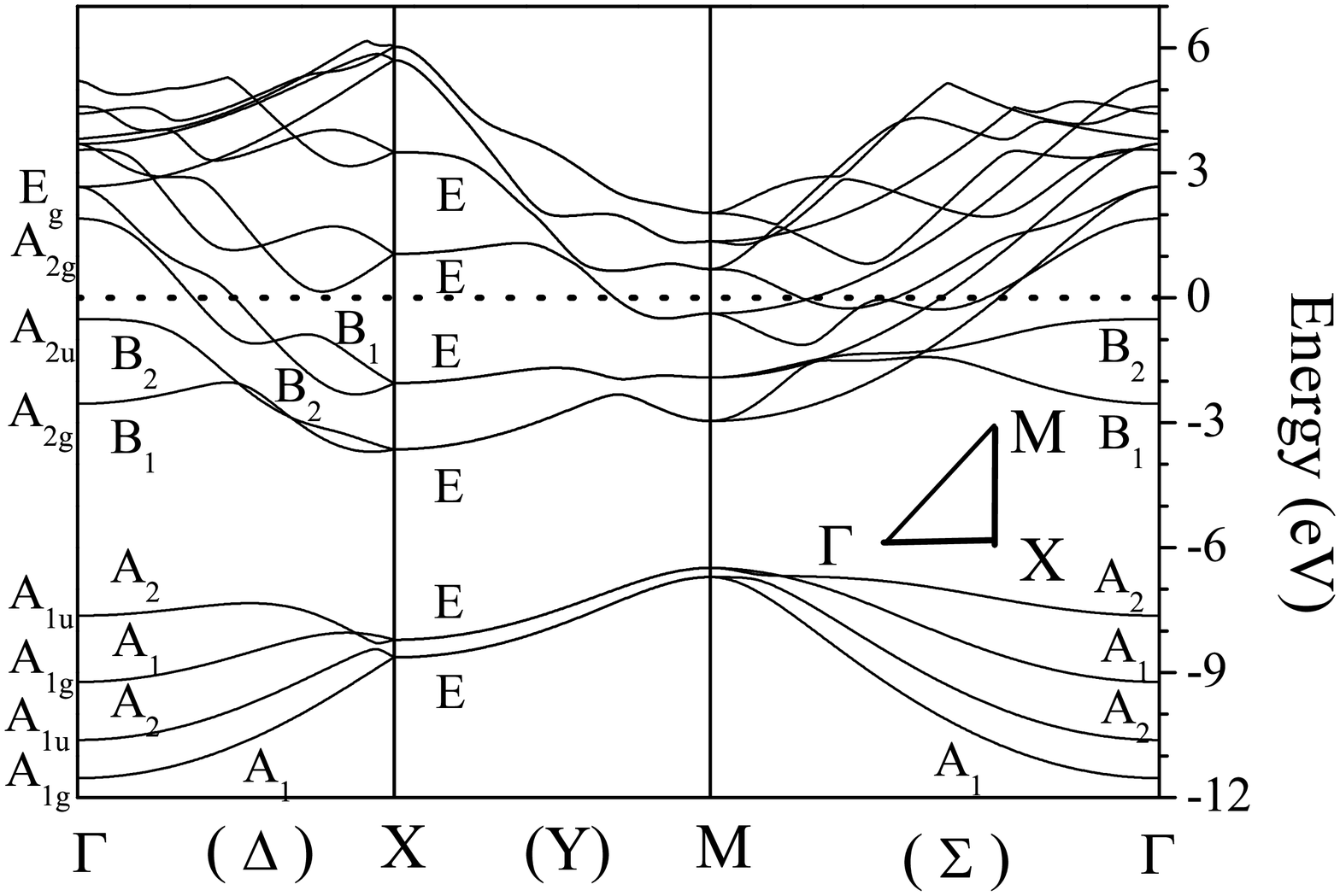}
\caption{Band structure of a four-layer, free-standing Pb(100)
film along high symmetry directions in the film Brillouin zone
(FBZ). These directions are depicted in the irreducible part of
the FBZ. The symmetry characters of the various states at the
$\Gamma$ point and along the symmetry lines in the FBZ are
indicated. Note that the states along the Y line are
two-fold degenerate in an even-layer film, and are nondegenerate
in an odd-layer film. Fermi energy is set to 0 eV.
}
\end{figure}

As a first step to understand the calculated QSE period, we have
studied the QW states of the isolated 2 to 12 layer thick Pb(100)
films. In Fig. 2 we show, as an illustrative example, the band
structure of a bulk-truncated, four-layer Pb(100) film in the
irreducible part of the film Brillouin zone (FBZ). The symmetry
characters of the bands are indicated.  For each wave vector in
the FBZ, the discrete set of bands corresponds to QW states. Note
that the bulk $s$ and $p$ subbands are separated by a gap of about
4 eV. The number of QW ($s$-QW) states derived from the bulk $s$
bands is given by the number of layers ($N$) in the film, whereas
there are $3N$ $p$-QW states derived from the  bulk $p$ bands.
Only $p$-QW states cross the Fermi level and are relevant for the
QSE. The $p$-QW states disperse either to higher energy or lower
energy away from the $\Gamma$ point depending on their orbital
character being $p_z(\sigma)$- or $p_{x,y}(\pi)$-like. These
states exhibit many ``avoided crossings'' accompanied with
interchange of their characters. A key observation about QW states
of these films that cannot be understood from the bulk band
structure is the lifting by odd-layer films of the two-fold
degeneracy of QW states along the Y symmetry axis in even-layer
films. Note that the projected bulk band structure on this axis is
two-fold degenerate.

\begin{figure}
\includegraphics[width=8cm]{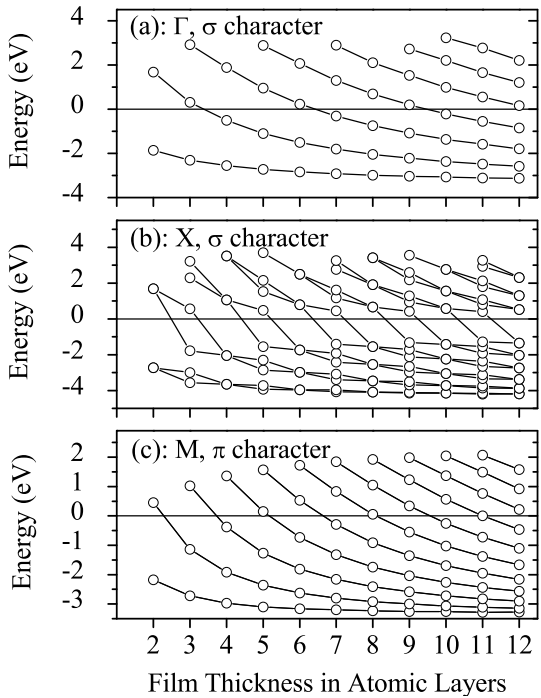}
\caption{Energy spectra of quantum well states derived from bulk p
bands at the high symmetry points $\Gamma$, X and M in the film
Brillouin zone  as a function of the number of atomic layers in
the film. At the $\Gamma$ and X points only states with $\sigma$
bonding character are shown, whereas at the M point only states
with $\pi$ bonding character are shown. The states with different
bonding characters at these {\bf k} points do not cross the Fermi
level and are not shown. Note that the splitting of the states
around the Fermi level in panel (b) exhibit an odd-even-layer
alternation. The Fermi energy is set to 0 eV.}
\end{figure}

To understand the effect of the odd-even layer alternation of the
degeneracy on the consecutive filling of the QW states, we show in
Fig. 3 the calculated layer dependence of the energy spectra of
$p$-QW states at high symmetry points in the FBZ. The $p$-QW
states of $\pi$ bonding character at $\Gamma$ and X points and
$p$-QW states of $\sigma$ bonding character at the M point do not
cross the Fermi level and are accordingly not
shown~\cite{footnotesigmapi}. The occupied bands at $\Gamma$ can
not accommodate all the $p$-electrons. In principle, it is
necessary to consider the QW states at the whole FBZ, in
particular at all the high symmetry points.

Only the QW states at the X  point have a bilayer
periodicity in Fig. 3. It is clear that the QW states at the
$\Gamma$ point cross the Fermi level by a period of about 3 layers in
good agreement with the perpendicular spanning vector in the
$\rm \Gamma X$  direction.  Note that at the M point, the
two-fold degeneracy of the $\pi$-like QW states are not lifted by
odd-layer films. However, at the X point (and along Y symmetry axis)
this degeneracy is lifted by odd-layer films corresponding to a bilayer periodicity.
In particular for states crossing the Fermi level, the splitting
is anomaly large because of a large gap of about 1.5 eV in
the projected bulk $p$ band structure along the Y symmetry axis.
The zone boundary band-gap
in the thin film therefore exhibits an odd-even-layer oscillation,
similar to that found in Si(100) thin films~\cite{Zunger}.

In general, QW states from different regions of the FBZ all
contribute to the film total energy. In particular, the QSE period
will be determined from the extremal perpendicular spanning
(nesting) vectors which determine the period of the enhanced DOS
at Fermi-level. In case that QW states at zone center dominate,
free-electron-like model is able to give reasonable prediction of
the film stability, e.g., Pb(111) film\cite{Wei}, Ag(100) film
\cite{ChiangS}. In the case of the Pb(100) thin film, the QW
states at BZ boundary give rise to the periodicity of the film
energy.

The different degeneracies along the FBZ boundary in odd and even
layer Pb(100) films is due to the symmetry of the layer stacking.
Because the Pb(100) film (and the (100) film of any fcc or bcc
metal) has an ABAB... stacking sequence, odd- and even-layer films
have different symmetry. An odd-layer film has an mirror plane on
the central layer. The point group is $ D_{4h}$, and the space
group is symmorphic. For an even-layer film, there is no such
direct mirror plane but a glide plane. The resulting space-group $
D_{4h}^7$ is non-symmorphic. For the odd-layer film, the point
group along Y axis is $C_{2v}$, and the bands of odd-layer film
are always non-degenerate along Y. For the even-layer films,
non-primitive translations are involved along this axis and a
multiplier representation has to be considered ( see e.g.,
Ref.~\cite{Burns}). Only a single two-dimensional representation
is found and the bands in even-layer films are all two-fold
degenerate along the Y axis.

So far the observed and calculated QW states of thin metal films has
been successfully described by the bulk band structure and the
phase accumulation model. However, as illustrated by the results for
the QW states of thin Pb(100) films, the energies of the QW states at the FBZ
boundary of (100) film of any fcc (or bcc) metal are determined
not only by the film thickness, but also by the symmetry of the
layer stacking geometry. A mirror plane for odd-layer film and a
glide plane for even-layer film result in non-degenerate and
two-fold degenerate QW states along Y axis. This symmetry effect can not
simply be captured by the bulk band structure and the phase
accumulation model. However, the size of the splitting across the
Fermi level and the QSE in film  energy and relaxation is material specific because
it depends on the presence of a large band gap at
the Fermi level at the boundary of FBZ. For instance, in the case of
a Al(100) film, we have a similar symmetry effect but this band-gap is
well above Fermi-level and the film energy QSE period is given by
the free electron result.

An important remaining issue is whether the bilayer periodicity of the
freestanding Pb(100) film energy survives on adsorption on a
semiconductor.  A simple estimate of the energy of a thin
Pb(100) film grown on a semiconductor surface using the
``electronic growth'' model~\cite{Zhang} indicates that the bilayer
periodicity survives. In this model, the total energy $E_t$ of a metal film
grown on a semiconductor substrate is given by $E_t=E_0-E_c$,
where $E_0$ is the (relaxed) film energy shown in Fig.1 and $E_c$
is the energy gain per surface unit cell due to charge transfer at
the metal semiconductor interface. This energy gain can be
estimated in a simple capacitor model:
$E_c=C(\Delta\Phi)^2$\cite{Zhang,Yeh}, where $C$ is a
phenomenological parameter specific to the interface and $\Delta
\Phi$ is the difference between the work functions of the clean
substrate and the film. For example, $E_c$ can be estimated in the
case of Si(111)-$(7\times7)$ semiconductor substrate by using the
same value for $C=0.033/(eV{\AA}^2)$  as used by Yeh {\sl et
al.} for similar Pb(111) films (similar charge density, work
function, surface energy, etc), and using the measured value 4.75
eV for the substrate $\Phi$~\cite{Yeh} and our calculated values
for the film $\Phi$. Using this estimate for $E_c$ and the
calculated values for $E_0$, the resulting values for $E_t$ are
{\bf 0.47}, 0.65, {\bf 0.54}, 0.57, {\bf 0.54}, 0.59, {\bf 0.54}
eV for {\bf 2}, 3, {\bf 4}, 5, {\bf 6}, 7 and {\bf 8} layers of
the Pb(100) film. Because the oscillations of the work functions
are in antiphase relative to those of $E_0$, $E_c$ partly
compensates the strong oscillation of $E_0$ but its odd-even layer
alternation has not been destroyed.

The above analysis does not necessarily suggest that Pb(100) films
can grow on Si(111)-($7\times7$). But it does illustrate that the
total energy can have an odd-even oscillation when a Pb(100) thin
film is grown on a semiconductor substrate. However, one has to
choose a substrate with a good lattice match so that the elastic
energy is negligible.  The Pb(100) film should then exhibit
``electronic growth" if care is taken to relieve possible
kinetic limitations.

Finally, we would like to discuss the nature of the quantization
of the QW states as obtained from a scrutiny of their wave
functions (WF). It is found that the $\sigma$- and the $\pi$-like
QW states form two independent series of
states~\cite{footnotesigmapi}.   As expected, the lowest QW state
derived from the bulk $s$-band has zero nodes, while the second
one has one node, etc. The bulk $s$-band accommodates 20 QW states
for a 20 layer film with a maximum of 19 nodes. The low-lying
$p_z$-QW states have $\sigma$ character. The lowest
 $p_z$-QW state  has 20 nodes and
 constitutes the next state in the $\sigma$ series,
followed by the $p_z$-QW states having 21, 22, 23, 24 nodes, etc. The
higher-lying $p_{x,y}$-QW states with $\pi$ character are two-fold
degenerate and form an independent series of QW states from the
$\sigma$ series of $s$- and $p_z$-QW states. The lowest-lying
$\pi$ state ($E_g$) has quantum number one (zero nodes), the
second $\pi$ level ($E_u$) has quantum number two (one node), etc.
The separate quantizations of $\sigma$ and $\pi$ states are rather
general. We have investigated the QW states in linear, isolated
chains of Na, Mg, Al, and Pb. The QW states spectra in these
linear chains are quite similar to that of the Pb(100) film at
$\Gamma$ point. The two-fold degenerate $\pi$ states are  higher
in energy and quantize independently from the $\sigma$ states.

In summary, we have shown from density functional calculations
that Pb(100) thin films exhibit quantum size effect with an
odd-even-layer alternation of film energies, film relaxations,
etc, which are due to the different symmetry of the stacking
geometry in odd and even layer Pb(100) thin films. The quantum
well states at the film Brillouin zone boundary have a two-fold
degeneracy in even-layer films, which are lifted in odd-layer
films. Furthermore, a large band-gap at the Brillouin zone
boundary in Pb(100) films makes the odd-even alternation of the
splitting of the bands embracing the Fermi level significant for
the film energy. This alternation of the film energy is argued to
survive on a semiconductor substrate. Even-layer Pb(100) films
should  be especially stable and may show up as ``magic
thickness''. Furthermore, we show that the quantum well states in
a simple-metal film can be classified into $\sigma$ and $\pi$
bonded states in the perpendicular direction of the film, which
quantize independently.


\begin{thebibliography}{99}
\bibitem{footnotesigmapi} {A $\sigma$-bonded state has no nodal plane along the bonding direction. A $\pi$-bonded state has one nodal
plane along the bonding direction.}
\end{thebibliography}
\end{document}